\documentclass{article}

\usepackage{arxiv}

\usepackage[utf8]{inputenc} 
\usepackage[T1]{fontenc}    
\usepackage{url}
\usepackage{amsmath}
\usepackage{amsthm}
\usepackage{mathtools, nccmath}
\usepackage{booktabs} 
\usepackage[utf8]{inputenc}
\usepackage[english]{babel}
\usepackage[T1]{fontenc}
\usepackage{textcomp}
\usepackage{graphicx}
\usepackage{color}
\usepackage{xspace}
\usepackage{subfigure}
\usepackage{multirow}
\usepackage{float}
\usepackage{epstopdf}
\usepackage{balance}
\usepackage{microtype}
\usepackage{hyperref}
\usepackage{listings}
\newtheorem{theorem}{Theorem}[section]
\newtheorem{corollary}{Corollary}[theorem]
\newtheorem{lemma}[theorem]{\textbf{Lemma}}
\usepackage{etoolbox}

\usepackage{array}
\newcolumntype{L}[1]{>{\raggedright\let\newline\\\arraybackslash\hspace{0pt}}m{#1}}
\newcolumntype{C}[1]{>{\centering\let\newline\\\arraybackslash\hspace{0pt}}m{#1}}
\newcolumntype{R}[1]{>{\raggedleft\let\newline\\\arraybackslash\hspace{0pt}}m{#1}}

\usepackage{url}
\usepackage{amsfonts}
\usepackage{times}

\AtBeginDocument{%
  \providecommand\BibTeX{{%
    \normalfont B\kern-0.5em{\scshape i\kern-0.25em b}\kern-0.8em\TeX}}}

\usepackage{cleveref}

\begin{document}






%

\title{The Hidden cost of the Edge:  A Performance Comparison of Edge and Cloud Latencies}

\author{
 Ahmed Ali-Eldin \\
  Chalmers University of Technology\\
  \texttt{ahmed.hassan@chalmers.se} \\
   \And
 Bin Wang \\
  University of Massachusetts, Amherst\\
    \texttt{binwang@cs.umass.edu} \\
  \And
 Prashant Shenoy \\
  University of Massachusetts, Amherst\\
    \texttt{shenoy@cs.umass.edu} \\
 }

\maketitle

%
%






\begin{abstract}
Edge computing has emerged as a popular paradigm for running latency-sensitive applications due to its ability to offer lower network latencies to end-users. In this paper, we argue that despite its lower network latency, the resource-constrained nature of the edge can result in higher end-to-end latency, especially at higher utilizations, when compared to cloud data centers. We study this edge performance inversion problem through an analytic comparison of edge and cloud latencies and analyze conditions under which the edge can yield worse performance than the cloud. To verify our analytic results, we conduct a detailed experimental comparison of the edge and the cloud latencies using a realistic application and real cloud workloads. Both our analytical and experimental results show that  even at moderate utilizations, the edge queuing delays can offset the benefits of lower network latencies, and even result in performance inversion where running in the cloud would provide superior latencies.  We finally discuss  practical implications of our results and provide insights into how application designers and service providers should design edge applications and systems to avoid these pitfalls.
\end{abstract}
\section{Introduction}
\label{sec:intro}

Over the past decade, cloud computing has emerged as a popular paradigm for running a variety of distributed systems applications ranging from web applications to Al and high performance computing workloads. In recent years, edge computing has emerged as a complement to cloud computing, particularly for running latency-sensitive workloads. Edge computing involves using computational and storage resources deployed at the edge of the network to run applications that can benefit from low network latency. Emerging edge applications include mobile augmented reality,  Internet of Things (IoT) analytics, and edge Al involving real-time inference over machine learning models. With the emergence of these edge workloads, many cloud providers have begun to offer edge cloud services by deploying edge servers clusters close to end users (akin to cloudlets \cite{satyanarayanan2009case})  and offering cloud-like service from the edge. 

Since edge resources are deployed at the edge of the network, conventional wisdom holds that 
the edge is ``better'' than the cloud from a latency perspective, and thus well suited for any workload that has tight latency requirements. In this paper, we show that this conventional wisdom does not always hold. In particular, edge applications  typically require low {\em end-to-end} latency, which consists of both the network and the server latency. While the edge provides a significantly lower network latency than the cloud, edge resources tend to be more constrained than those in cloud data centers. Consequently, workload dynamics can cause edge server latency to exceed the cloud server latency due to higher queuing delays at edge sites. In such cases, the total end-to end  latency of the edge can {\em exceed}  the end-to-end cloud latency, since the lower network latency of the edge is offset by a higher server latency at higher utilization levels. This "hidden" cost of the edge has not been studied previously. Our paper quantifies this cost through analytic and experimental comparisons of edge and cloud performance. Specifically, we use analytic queuing results and real world experimentation to characterize scenarios under which such performance inversion occurs. 

In doing so, our performance comparison study makes the following contributions
\begin{enumerate}
    \item  We introduce and formulate the {\em edge performance inversion} problem,  which causes the end-to-end latency of the edge to become higher  than that of the cloud. We conduct an analytic performance comparison by using queuing models to  to quantify the end-to-end latency of the edge and the cloud under different workload scenarios. We present closed-form analytic equations that specify conditions under which  the total edge latency can exceeds the cloud latency, causing a performance Inversion 
\item We conduct an experimental performance comparison of the edge and cloud performance
using realistic applications and Azure trace workloads  on  real cloud and edge platforms.   Our experiments show that for geo-distributed applications the mean and tail latencies of the edge can indeed be worse than the cloud even at moderate utilization levels of 40 to 60\%   and that such performance inversion is more likely in the presence of workload skews or as the latency to traditional clouds reduces due to increasing cloud deployments. 

\item Since the potential for performance inversion at the edge has important consequences for both  application designers and cloud providers,  we discuss the key implications for our results in practice.  Specifically, we discuss resource allocation techniques and extra capacity needed at the edge to avoid performance inversion with the cloud under different scenarios.


\end{enumerate}
 \section{Background}

This section presents background on  cloud 
 and edge computing discussing the  motivation for the edge performance inversion problem. 

 \subsection{Cloud and Edge Computing}
 Traditional cloud computing involves deploying large-scale data centers that  house tens of thousands of servers to run remote third-party applications. Customers can request these servers to run their applications and pay based on their resource usage on a pay-as-you-go basis. Cloud data centers are typically virtualized and allocate server resources to applications in the form of virtual machines. Today, cloud platforms are popular for running diverse applications ranging from online web services to scientific computing.
 
 Edge computing and edge clouds are a natural evolution of traditional cloud computing \cite{satyanarayanan2017emergence}. They involve deploying server resources at the edge of the network to host workloads that are latency-sensitive in nature.  While there are many forms of edge computing, in this work, we focus on edge clouds that involve deploying small server clusters, also known as micro data centers, at the geographically distributed sites to offer a cloud-like service from the network edge. 
 The notion of edge clouds was originally proposed as an extension to the work on pervasive  and mobile computing in the form of \emph{Cloudlets}~\cite{satyanarayanan2009case}. Major cloud providers are now beginning to offer edge cloud services (e.g., Google's Stadia \cite{stadia} and Azure IoT Edge service \cite{azure-edge}). Edge clouds are especially useful for applications that demand low-latency access to cloud resources which can not be satisfied using more distant traditional cloud servers.  
 
 

  Cloud workloads exhibit significant temporal  and spatial dynamics~\cite{khan2012workload,lu2017imbalance}, and we expect edge workloads to exhibit a similar behavior.  Temporal workload dynamics include diurnal time-of-day and seasonal effects as well 
 as short-term burstiness and workload spikes in the form of flash crowds. Since the end-users accessing a cloud or edge application may be geographically distributed, the workload also exhibits spatial dynamics. If users are more concentrated at certain locations than others, the incoming workload will exhibit a skew in its spatial distribution. Moreover if the popularity of the application decreases in certain regions and rises in others, the spatial workload distribution will change over time. From the perspective of a traditional cloud applications, all  requests arrive at a centralized data center and these temporal and spatial dynamics are handled at a single location via techniques such as dynamic resource allocation or elastic scaling~\cite{xue2018spatial}. However, in the case of a distributed edge cloud,  different edge sites will see different spatial and temporal dynamics since an edge  application such as an online game may be distributed across  geographic sites and thus the overall workload will  be partitioned across sites.  We assume that applications have tight end-to-end latency requirements that need be satisfied by the remote server, whether at the edge or the cloud. 
 
 \subsection{Edge Performance Inversion: Motivation}
 \label{sec:problem}

Next, we motivate the potential for edge performance inversion using several intuitive observations.
Consider an application that can either be deployed in the cloud or at the edge. Suppose that the roundtrip network latency to the 
cloud server is $n_{cloud}$ and that to the edge server is $n_{edge}$. 
Since edge resources are closer to end-users than cloud resources, we have $n_{edge}<n_{cloud}$. This lower network latency of edge servers has been the primary reason as to why edge computing 
has been assumed to be better than cloud computing for latency-sensitive applications. 

\begin{figure}
\centerline{
 \subfigure[Edge queuing model.] 
{\label{fig:distQ} \includegraphics[height=1.7 in]{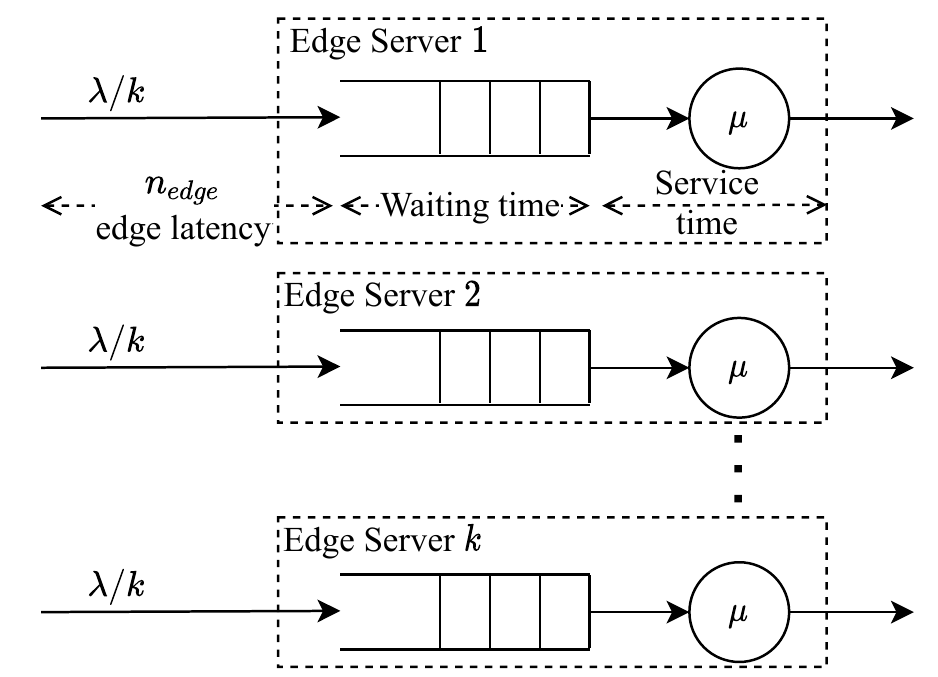}}
  \subfigure[Cloud queuing model.] 
{\label{fig:centQ} \includegraphics[height=1.7 in]{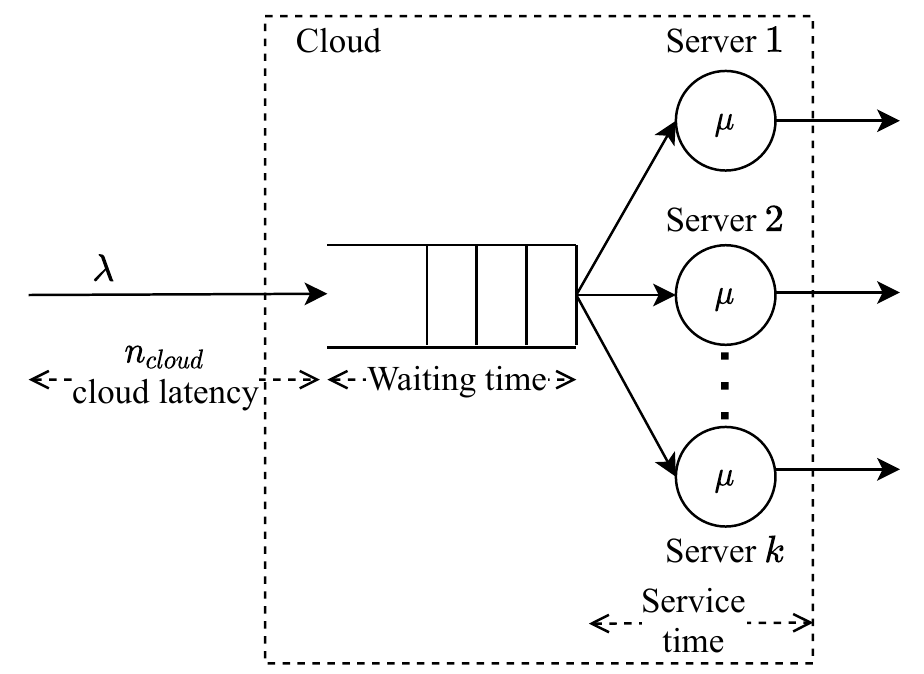}}
}
 \vspace{-14 pt}
 \caption{End-to-end latency of an edge and cloud deployment of an application. }
  \vspace{-0.5 cm}
\label{fig:qs}
 \end{figure}

From an application perspective, however, the performance is governed not just by network latency but  by  the total end-to-end latency.  Let $r_{edge}$ denote the response time of the edge server and $r_{cloud}$ denote the response time of the cloud server for each application request. Thus, the end-to-end latency offered by the edge is ($n_{edge}+r_{edge}$) while that of the cloud is ($n_{cloud}+ r_{cloud}$). As shown in Figure~\ref{fig:qs}, the server response 
time itself consists two components; queuing delays at the server 
and the request execution time.

\noindent\textbf{Bank teller analogy:} Suppose that the application runs on a cluster of $k$ servers
on the cloud.  It is common for many cloud-based applications to run on a cluster of servers.
 Let the cloud request rate seen by the application be $\lambda$ req/s. If this application were to be deployed at the edge instead of the cloud, we assume that it still uses $k$ servers, but that these servers are 
distributed across multiple edge sites. In the limit, the application can use $k$ edge sites, each hosting one 
server for the application. Suppose that the incoming workload of $\lambda$ req/s is  uniformly (equally) distributed across these $k$ edge sites. As shown in  Figure \ref{fig:qs}, each edge site sees a workload $\lambda/k$ req/s. 

In this case, the  server latencies at the edge and the cloud   can be viewed from the perspective of the well-known bank teller problem that compares a separate queue per teller versus a single queue for all tellers, a problem well studied in the queuing theory literature since the seventies
~\cite{kingman1970inequalities,rothkopf1987perspectives,whitt1992understanding}.

The main insight of the bank teller problem is that customers will always see lower waiting times when using  a single queue  for all tellers versus a separate queue per teller.  Since the time needed to service each customer  varies from customer to customer, in the latter case, some queues see longer waiting times when some customers from those queues make long transactions. 
A centralized  queue avoids such problems since there is a single queue and all queued customers see the same impact.  In line with this intuition, queuing theory shows that a centralized queue  always yields lower wait time with a few known exceptions (e.g., when jockeying between queues is permitted~\cite{ rothkopf1987perspectives}).

In our case, the centralized queue with $k$ tellers is analogous to the cloud deployment of the application with $k$ servers and a single arrival queue.  A separate queue per teller is analogous to each edge site with a single server that services its own queue.
Thus, even when the workload is equally distributed across edge sites, queuing theory from the bank teller problem tells us that $r_{edge}>r_{cloud}$, since waiting times at individual queues at the edge will be higher than that of the cloud. 
In fact, as at high utilization levels,  the queuing delays at the edge can be \emph{factor of k
times} higher than the cloud (see~\cite{harchol2013performance} for a proof).

Thus the edge has lower network latency than the  cloud $n_{edge} <n_{cloud}$, but has higher server latency than the cloud $r_{edge} > r_{cloud}$. If the higher queuing delays at the edge offset the lower network latency, the total end-to-end latency of the 
edge can actually become higher than that of the cloud. 
That is $(n_{edge}+ r_{edge}) > (n_{cloud} + r_{cloud})$. We refer to this problem as the {\em performance Inversion problem} of the edge.
Such performance inversion only occurs at higher utilization levels, where edge queuing delays are significantly higher than the cloud. Thus, it is important to analyze and quantify the utilization levels above which performance inversion will occur.  This problem can be posed as a variant of the bank teller problem with driving times. A customer can drive to a nearby local bank branch that has a single teller, or they can drive to the 
main bank branch, which is further away but has $k$ tellers and a single queue. The total time now includes driving time (i.e., network latency) and the time spent at the branch (server 
latency), which also includes waiting times in the queue. This variant of the bank teller problem with driving times has not been studied previously.

 \begin{figure} 
\begin{center}
\includegraphics[width=3in]{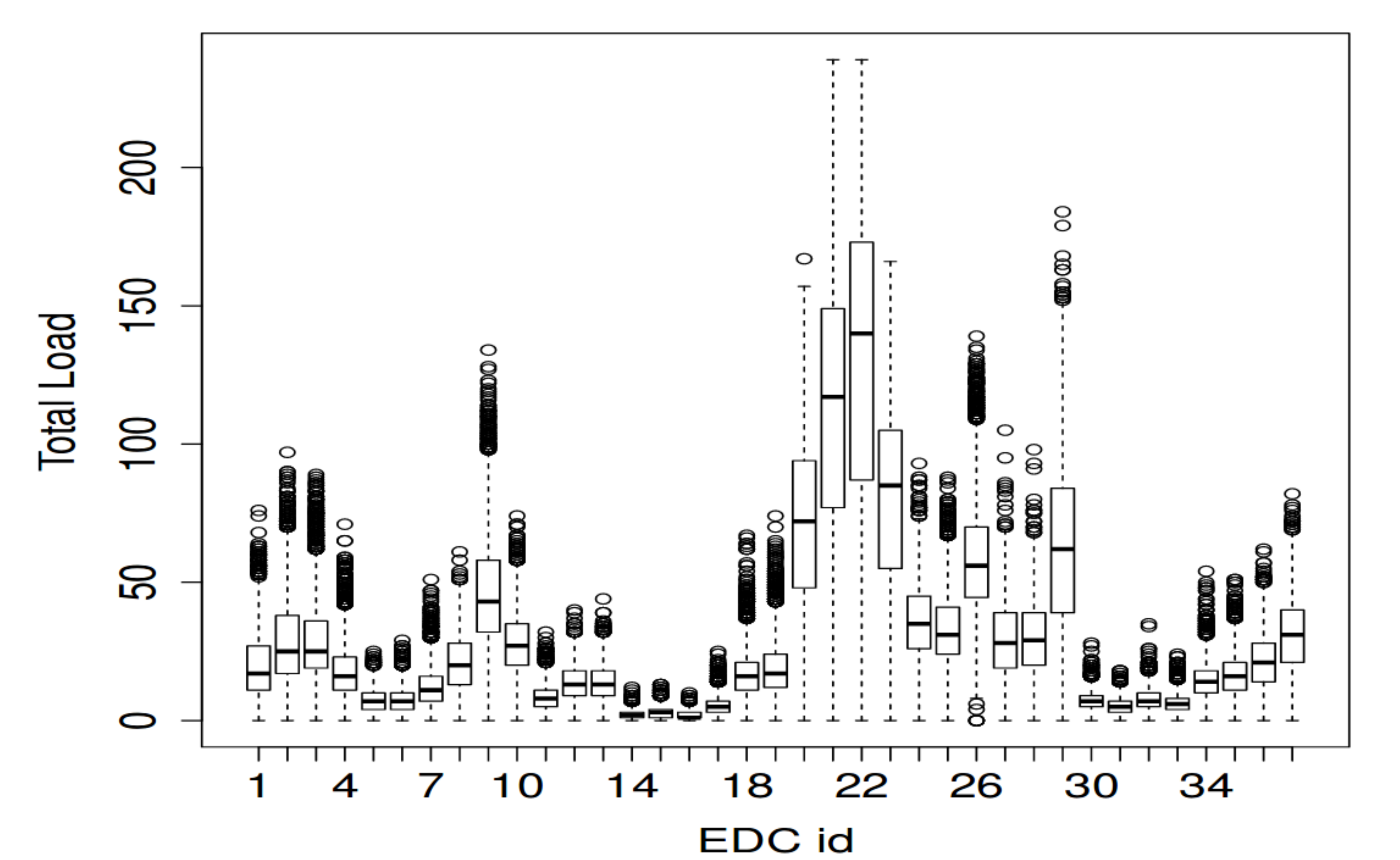}
\end{center}
\vspace{- 4 mm}
\caption{Non-uniform geographic distribution of the  total load, measured as the number of vehicles in a cell across time, seen by edge data centers for a vehicular edge application \cite{le2017location}.}
\label{fig:nu}
 \end{figure}

\noindent\textbf{Impact of workload skews.} Our bank teller analogy assumed that the aggregate workload $\lambda$ is equally partitioned across 
the $k$ edge sites, yielding a workload of $\lambda/k$ per edge site. Such balanced workload is the ideal case for the edge.  In practice, edge and cloud workloads will exhibit spatial and temporal dynamics that cause load skews. In particular some 
edge sites may see higher request rates than others due to spatial dynamics, causing the total workload 
to be split \emph{unequally} across sites. Spatial and temporal dynamics will mean that these imbalances will change over time and also cause the set of edge sites 
that see higher arrivals to change over time.

These workload skews exacerbate the performance 
inversion problem, since some edge sites will now see higher waiting times due to higher arrival rates at those sites. Workload skews can result in performance inversion even at lower utilization levels. Note that skews have little impact on the cloud since the cloud still sees the same workload of $\lambda$. 
Workload skews have been observed in empirical studies. For example, Le Tan et al.~\cite{le2017location} have shown that load on edge clouds is non-uniformly spatially distributed, using GPS traces of taxis in SanFrancisco~\cite{piorkowski2009crawdad}, and assuming each edge  cloud serves a hexagonal area of radius 1 km. Figure~\ref{fig:nu} shows a box-plot of the load seen on each cell based on the data from~\cite{piorkowski2009crawdad}, with some cells having significantly larger loads than others~\cite{le2017location}. In other words, while some edge  cloud sites will be highly utilized, others will experience low utilization as seen in Figure~\ref{fig:nu}. The load will shift between day and night based on humans' diurnal mobility patterns~\cite{gonzalez2008understanding}. 

The above observations motivates the two principal research questions addressed in this paper 
\begin{enumerate}
    \item Under what scenarios and utilization levels does the lower network latency of the 
edge get offset by higher queuing delays, causing a
performance inversion where total edge latency becomes  worse than the cloud latency? 
\item How do workload skews impact the queuing delays and utilization levels across edge sites and what is its impact on performance inversion at the edge? 
\end{enumerate}
These consequences of these questions for application developers and cloud platforms  is addressed through a third research question:  (3) How can an  application designer reduce the chances of performance inversion for their edge application, and how should a cloud provider ensure its edge cloud customers do not see 
performance inversion problems? 

\section{Analytic Performance Comparison}
\label{sec:latency-model}

As  discussed above, the edge offers lower network latency than the cloud but can impose higher queuing delays, and thus higher server latency, in some scenarios. In this section, we analytically model the edge and the cloud queuing delays to understand when performance inversion can occur at the edge due to these two opposing factors.

\subsection{Modeling Edge and Cloud Latency}

As noted in $\S$\ref{sec:problem},  we consider an application that is assumed to run on $k$ cloud servers, $k \geq 1$,  to collectively service a mean workload of $\lambda$ requests/s.  When the same application is deployed at the edge, the $k$ servers are distributed across up to $k$ geographic edge site locations and incoming requests are sent to the closest edge site for service.  For the ease of analytic modeling, we assume  that the application uses  $k$ edge locations, each hosting one server.  All of our results are easily extended to the general case where each each site hosts more than one server.  


Initially, we assume that the workload is balanced across all edge sites (we relax this assumption and consider the impact of workload skews in $\S$\ref{sec:skew}.)  Thus, each edge site receives 
$\lambda/k$ request/s.  Let $n_{edge}$ and $n_{cloud}$ denote the network round trip  latency to the edge and the cloud, respectively. Since edge servers are more proximal to end-users,  $n_{edge} < n_{cloud}$.   The server latency at the edge and the cloud consists of two components: the queuing delay experienced by a request and the service time to execute the request. 
Let $w_{edge}$ and $w_{cloud}$ denote the queuing delay (i.e., waiting time) incurred by a request upon arrival at the edge and the cloud server, respectively. Also, let $s_{edge}$ and $s_{cloud}$ denote the service time of the request at the edge and cloud, respectively. The end-to-end latency seen by a request is  the sum of the network latency, server queuing delay, and the request service time. That is,
\begin{equation}
    T_{edge}=n_{edge}+w_{edge}+s_{edge}
\end{equation}
and 
\begin{equation}
    T_{cloud}=n_{cloud}+w_{cloud}+s_{cloud}
\end{equation}
where $T_{edge}$ and $T_{cloud}$ denote the total end-to-end latency of the edge and the cloud, respectively.


\subsubsection{Edge and Cloud Latency Bound} 
\label{sec:MMSd}

Queuing theory is a useful mathematical tool for analyzing the waiting time (queuing delays) seen by a  requests at an edge or cloud server.  Queuing theory provides well-known closed-form 
results for waiting times under different workload arrival distributions and different  server scheduling disciplines \cite{harchol2013performance}. We can use these well-known results to  compare the queuing delays at the edge and the cloud at different utilization levels and derive closed-form equations to determine when higher queuing delays at the edge will cause a performance inversion. Such equations provide easy ``rules of thumb'' for a system designer or an application developer to analyze whether their edge applications are vulnerable to a performance inversion.

To perform such an analysis, we must model the edge and cloud servers as queuing systems,  as shown in Figure \ref{fig:qs}.  The simplest type of queuing model assumes that the server sees Poisson arrival of requests, and that request service times follow an exponential distribution.  In this case, the single server at each edge site is modeled as a $M/M/1$ queuing system, while  the $k$ servers at the cloud are collectively modeled as a $M/M/k$ queuing system.  These queuing models yield our first key result.

\begin{lemma}
The end-to-end latency of the edge  is higher than the cloud whenever the network round trip latency difference between the edge and the cloud $\Delta n$ is less than $\sqrt{2}\left( \frac{1}{1-\rho_{edge}} -\frac{1}{\sqrt{k}(1-\rho_{cloud})}\right)$
\label{lemma:mmk}
\end{lemma}
\noindent\textbf{Proof:} 
The edge  will offers worse end to end latency than the  cloud when $T_{edge} > T_{cloud}$. That is
\begin{equation}
    n_{edge}+w_{edge}+s_{edge} > n_{cloud}+w_{cloud}+s_{cloud} \label{eq:cond}
\end{equation}
Assuming that both the edge and  cloud employ the same hardware configuration for their servers
and incoming requests are serviced using the FCFS service discipline, the execution time of a request on an edge server and the cloud server will be identical.
That is, $s_{edge} = s_{cloud}$. The above inequality then reduces to
\begin{equation}
    n_{edge}+w_{edge}  > n_{cloud}+w_{cloud}   
\end{equation}
Let $\Delta n$ denote the difference in network round trip times between the edge and cloud data centers. That is 
$\Delta n =  n_{cloud} - n_{edge}$. Substituting $\Delta n$ in the above inequality yields
\begin{equation}
   \Delta n <  w_{edge} - w_{cloud}   \label{eq:deltan}
\end{equation}
This inequality formally states our intuition---that the edge will  worse than the cloud when the reduction in network latency at the edge if offset by higher queuing delay at the edge than the cloud. 

Our bank teller analogy intuitively tells us that the single queue for $k$ servers at the cloud should yield an overall lower queuing delay at the cloud than that at the edge. We now formally 
analyze the queuing delays offered by the  edge and the cloud.  To do so, we use  a well-known queuing theory result from Whitt et. al~\cite{whitt1992understanding}, which shows that for the same probability of delay, a system using a multi-server queue requires less resources than one using multiple single server queues.  
Further, the analysis in \cite{whitt1992understanding} also states that the expected waiting time for a request in a system of $k$ servers is
\begin{equation}
    E[w|w>0] =\frac{\sqrt{2}}{(1-\rho)\sqrt{k}}. \label{eq:ExpWM}
\end{equation}

Applying this result to Equation \ref{eq:deltan}, we have
\begin{equation}
   \Delta n <  E[w_{edge}|w>0] - E[w_{cloud}|w>0]   \label{eq:deltan2}
\end{equation}
where $E[w_{edge}|w>0]$ and $E[w_{rem}|w>0]$ are the expected queuing delays for the edge and the cloud system, given that the queuing delay is non-zero. We note that the accuracy of Whitt's approximation result in Equation \ref{eq:ExpWM} increases with higher utilization, since queuing delays are more likely to be non-zero at high utilization levels. 
Substituting  Equation \ref{eq:ExpWM}, we get 
\begin{equation}
   \Delta n <  \frac{\sqrt{2}}{(1-\rho_{edge})}  -  \frac{\sqrt{2}}{(1-\rho_{cloud})\sqrt{k}}  
\end{equation}
which completes the proof.    \hfill \qed

We now explain the practical implications of the above result by deriving several corollaries and discussing the key takeaways.
\begin{corollary} \label{corr:equal_balanced}
In the case where the application workload is equally balanced across edge sites and the same server configuration is used by the edge and the cloud, the above result reduces to 	
$\Delta n < \frac{\sqrt{2}}{(1-\rho_{edge})} (1 - \frac{1}{\sqrt{k}})$, which yields the a bound on the server utilization above which edge performance inversion occurs
\begin{equation}
	\rho_{edge} > 1 - \frac{2}{\Delta n} (1 - \frac{1}{\sqrt{k}})
\end{equation}
\label{corr:rho}
\end{corollary}
\emph{Proof:}
In any queuing system, the utilization $\rho$ is given by the ratio of the request arrival rate and the request service rate (i.e., departure rate or system throughput). In case of the edge, each edge site sees an arrival rate of $\lambda / k$. The request execution time is $s_{edge}$, which means that the server can process $\mu = 1 / s_{edge}$ requests/s.   Hence $\rho_{edge} = (\lambda / k) / \mu = \lambda / k\mu$.
In case of the cloud, the arrival rate is $\lambda$. The execution time of a request is same as that of the edge server since the hardware configurations are assumed to be the same. Hence the service rate of a server is $\mu$, same as the edge server. Since there are $k$ cloud servers, the total service rate is $k\mu$. Hence, $\rho_{cloud} = \lambda / k\mu$    Since $\rho_{edge}$ is same as $\rho_{cloud}$, we can substitute $\rho_{edge}$ for $\rho_{cloud}$ in the above lemma, and the result reduces to 
\begin{equation}
\Delta n < \frac{\sqrt{2}}{(1-\rho_{edge})} (1 - \frac{1}{\sqrt{k}})
\end{equation}
Rearranging $\rho_{edge}$ in this inequality, we get 
\begin{equation}
	\rho_{edge} > 1 - \frac{2}{\Delta n} (1 - \frac{1}{\sqrt{k}})
\end{equation}   \hfill \qed

\begin{corollary} \label{corr:large-k}
As the number of edge locations $k$ increases, the cutoff edge  utilization that yields a performance inversion becomes a function of only   $\Delta n$. 
\end{corollary}

\emph{Proof:} 
As $k\rightarrow \infty$, the term $\frac{1}{\sqrt{k}} \rightarrow 0$ in the above inequality, which yields
$\rho_{edge} > 1 - \frac{2}{\Delta n}$  \hfill \qed 
\vspace*{0.1in}

\noindent {\bf Practical takeaways:}  The above corollaries provide the system designer with a cutoff utilization threshold above which edge performance inversion will occur.  Given the choice of an edge deployment that offers network latency $n_{edge}$ or a cloud deployment that offers a network latency $n_{cloud}$, the above results offers  simple rules of thumb for comparing these two deployments---if the expected system utilization will  be lower than $1 - \frac{2}{\Delta n} (1 - \frac{1}{\sqrt{k}})$, the edge will indeed provide a lower end-to-end latency. Otherwise the cloud offers a better choice.  Further, for a more geographically distributed edge deployment with large $k$, the utilization needs to be lower than $1 - \frac{2}{\Delta n}$ to avoid a performance inversion. 

It is also important to understand the applicability of these results in practice. Our analysis assumed a  \emph{multi-server application} that either runs on $k$ servers in the cloud or on a set of  geo-distributed servers at $k$ edge sites, where $k>1$.  Consider the special case where $k=1$, i.e., an application that run at a  \emph{single edge site} or one that runs on a single server (which also implies it runs at a single edge site). In either case, edge performance inversion will \emph{never occur}. To see why, when the entire application runs at a single edge site (i.e., is not geo-distributed), the entire cloud workload of the application is also seen by this  one edge site.  Since the server configuration is identical in both cases, the server latencies at the cloud and the edge are also identical. Hence the edge will always provide better end-to-end response time due to its lower network latency. This can be seen in Corollary \ref{corr:rho} where substituting $k=1$, reduces the requirement for performance inversion to $\rho_{edge} > 1$, which can never be true since utilization can not exceed 1. 
Interestingly, performance inversion can still occur for the case of $k=1$ if the edge uses a {\em different server configuration} than the cloud, and specifically if the edge uses more resource-constrained servers (e.g., servers with fewer cores or slower processors). In this case, the edge request execution times are slower than those at the cloud. This results in higher queuing delays at the edge than the cloud, implying that performance inversion can still occur at the edge for the case of $k=1$. 

The impact of using more resource-constrained servers at the edge also makes performance inversion more likely for multi-server geo-distributed applications (i.e., when $k>1$). In this case, the execution time of requests are no longer the same at the edge and the cloud, and  $s_{edge} > s_{cloud}$. Further performance inversion will occur if  $\Delta n <  (w_{edge} - w_{cloud}) + (s_{edge} + s_{cloud})$.  Since   
$s_{edge} + s_{cloud} > 0$, the right side of the inequality is larger than our result in Lemma \ref{lemma:mmk}, implying that performance inversion becomes  more likely.

\begin{corollary} 
\label{corr:lb}
A hard lower bound on the cloud network latency below which the edge  yield worse response time than the  cloud is given by 
\begin{equation}
    \sqrt{2}\left( \frac{1}{1-\rho_{edge}} -\frac{1}{\sqrt{k}(1-\rho_{cloud})}\right) \label{eq:Tau_rho_zeroedge},
\end{equation}
\end{corollary}

\emph{Proof:}  The best case for the edge network latency is for a very proximal edge where 
 $n_{edge}$ is nearly equal to zero.  Substituting $n_{edge}=0$ in $\Delta n$, 
the  lemma reduces to 
\begin{equation}
    n_{cloud}< \sqrt{2}\left( \frac{1}{1-\rho_{edge}} -\frac{1}{\sqrt{k}(1-\rho_{cloud})}\right).
\end{equation}
We can see that if $n_{cloud}$ is lower than the expression on the right, the lemma is always true  and the edge will always yield worse end to end latency than the cloud. 
 Hence, the expression  is a hard lower bound on the cloud network latency, below which the edge yields worse performance than the cloud.  \hfill \qed

\noindent \textbf{Practical takeaways:} Since major clouds platforms such as Amazon EC2 and Azure have begun to deploy additional  data centers in various geographic regions, the network latency to the  cloud 
 is decreasing when applications choose the closest cloud data center.
  Corollary~\ref{corr:lb} implies that if the cloud network latency drops below a certain threshold through deployment of regional data centers, it has the potential (at certain utilization levels) to offer overall end to end response times that are ``good enough'' for edge applications and lower than what a smaller edge site can provide. Put another way, with increasing  regional cloud data deployments, the benefits of the lower network latency offered even by the best edge deployments will diminish and the cutoff utilization level for performance inversion will also be lower, allowing the cloud to be "good enough" even at lower utilization levels. 
  


\subsubsection{Generalized Latency Bounds}\label{GGSd}
Our analysis so far assumed that application request arrivals and service times follow the Poisson and exponential distribution,  respectively. This assumption was made for analytical tractability, but real application workloads can have any arbitrary arrival and service time distributions. We now extend the above analysis to the general case.  To do so, we assume each edge site is modeled
as a G/G/1 queuing system and the cloud is be modeled as a G/G/k system. In queuing theoretic terminology, G/G/k  refers to a queuing system of $k$ servers that service a workload with 
with a general (i.e., arbitrary) distribution for arrival rates and service times.

\begin{lemma}
Assuming that the edge and the cloud are modeled as G/G/1 and G/G/k queuing systems, respectively,
the edge  offers higher  end-to-end latency times whenever the network latency difference between the cloud and the edge 
$\Delta n$ is less than
\begin{equation}
\begin{split}
&\rho_{edge} \frac{1}{\mu_{edge}(1-\rho_{edge})}\frac{c_{edge_A}^2+c_{edge_B}^2}{2} \notag \\ 
& \qquad -\frac{\rho_{cloud}^k +\rho_{cloud}}{2} \frac{1}{\mu_{edge}(1-\rho_{cloud})}\frac{c_{cloud_A}^2+c_{edge_B}^2}{2k}.
\end{split}
\end{equation}
\end{lemma}
\noindent \textbf{Proof:} Although there are no closed-form equations for  waiting times (queuing delays) in a G/G/1 and G/G/k queuing systems, there are several good approximations for waiting times that yield closed-form equations.  One such approximation for the expected waiting time is the Allen-Cunneen approximation~\cite{bolch2006queueing,KELLYBOOTLE1990247}, a widely used queuing theory result 
that has been shown to be reasonably accurate at higher utilization levels~\cite{whitt1993approximations}, and has found many practical applications~\cite{ahmad2010joint,hong2011dynamic}.


The Allen-Cunneen approximation states that the expected waiting time for a G/G/1 queue is,
\begin{equation}
   E(w) \approx \frac{\rho}{\mu(1-\rho)}.\frac{c^2_A+c^2_B}{2}, \label{eq:Allen1}  
\end{equation}
where, $c_A^2$ and $c_B^2$ are the squared Coefficients of Variation (CoV) of the inter-arrival time and the service times respectively. 
For G/G/k systems ~\cite{bolch2006queueing, whitt1993approximations}, the approximation for the expected waiting time  is given as
\begin{equation}
    E(w) \approx \frac{P_s}{\mu(1-\rho)}.\frac{c^2_A+c^2_B}{2k}, \label{eq:Allen2}  
\end{equation}
where $P_s$ is the steady state probability that an arriving request has to wait in the queue for a server to become available. Bolch et al.~\cite{bolch2006queueing} have shown that $P_s$ can be approximated closely as
\begin{equation}
    P_s \approx \begin{cases}
    \frac{\rho^k +\rho}{2} , & \text{if}\  \rho>0.7. \label{eq:Bosch1} \\
    \rho^\frac{s+1}{2},  & \text{if}\  \rho<0.7    
    \end{cases}
\end{equation}

Since the Allen-Cunneen approximation is more accurate in higher utilization regimes, and since edge performance inversion is likely at higher utilization levels, we consider the higher utilization case from 
Equation~\ref{eq:Bosch1}, i.e., $\rho>0.7$. 
Since $\Delta n = w_{edge} - w_{cloud}$ as noted in  Equation~\ref{eq:cond}, substituting Equation~\ref{eq:Allen1}, Equation~\ref{eq:Allen2} and Equation~\ref{eq:Bosch1} in Equation~\ref{eq:cond}, we get
 \begin{equation}
 \begin{split}
    &\Delta n < \rho_{edge} \frac{1}{\mu_{edge}(1-\rho_{edge})}\frac{c_{edge_A}^2+c_{edge_B}^2}{2} \\ & \quad - 
        \frac{\rho_{cloud}^k +\rho_{cloud}}{2} \frac{1}{\mu_{cloud}(1-\rho_{cloud})}\frac{c_{cloud_a}^2+c_{cloud_B}^2}{2s} 
        \end{split}
        \label{eq:ggsc}
\end{equation}
 where, $c_{cloud_A}^2$, $c_{cloud_B}^2$, $c_{edge_A}^2$ and $c_{edge_B}^2$ are the squared CoVs of the inter-arrival times and the service times of requests for the  cloud and the edge , respectively. Like before, we assume that the edge and the cloud use the same hardware configuration for servers
 and FCFS service disciplines. Hence the service times of requests on the edge and cloud are the same. Since service times are inverse of the service rate, we have $s_{cloud} = s_{edge} = 1/\mu_{cloud} = 1/\mu_{edge}$. For the same reason, the CoV of service times on the edge and cloud servers are the same: $c_{cloud_B}^2=c_{edge_B}^2$. Equation \ref{eq:ggsc} then simplifies to

\begin{equation}
\begin{split}
    &\Delta n <
    \rho_{edge} \frac{1}{\mu_{edge}(1-\rho_{edge})}\frac{c_{edge_A}^2+c_{edge_B}^2}{2}
    \\ & \quad -
    \frac{\rho_{cloud}^k +\rho_{cloud}}{2} \frac{1}{\mu_{edge}(1-\rho_{cloud})}\frac{c_{cloud_A}^2+c_{edge_B}^2}{2k}, \label{eq:ggsT}
    \end{split}
\end{equation}  \hfill \qed

\begin{corollary}
\label{corr:GGI}
 As the number of edge locations $k$ increases  $\Delta n$ becomes solely a function of the edge workload parameters
\begin{equation}
\begin{aligned}
    \Delta n < 
    \rho_{edge} \frac{1}{\mu_{edge}(1-\rho_{edge})}\frac{c_{edge_A}^2+c_{edge_B}^2}{2}\label{eq:ggsS}
\end{aligned}
\end{equation}

\end{corollary}

\emph{Proof:} As $k\rightarrow \infty$, the term  $\frac{c_{cloud_A}^2+c_{edge_B}^2}{2k}$ $\rightarrow 0$, and so does the second term of the inequality in Equation~\ref{eq:ggsT}, yielding the above inequality. \hfill \qed

\noindent\textbf{Practical takeaways} 
Unlike our M/M/k results which depended solely on the server utilization levels, our general results indicates that edge performance inversion depends both on the utilization levels as well as the bustiness of the workload , which is captured by the coefficient of variation (CoV) of arrival rates and service times.  

Corollary~\ref{corr:GGI}  implies that spikes, flash crowds, and high variability in the edge workload inter-arrival (and service) times can have a significant  impact on whether an edge will see a performance inversion. When the workload is bursty in nature, the CoV of the inter-arrival times will be highly variable. This makes performance inversion more likely and  implies that edge  is less suitable for applications where the request rates (and hence the inter-arrival times) have higher bustiness.


\subsection{Impact of Workload Skews}
\label{sec:skew}
 Our analysis thus far assumed equally partitioned  edge workload across edge sites. However, application workloads are unlikely to be geographically balanced and exhibit spatial 
skews (as shown in Figure~\ref{fig:nu}). For example, an edge application such as an online game may be more popular in certain 
parts of the world than others causing the aggregate workload to be split unequally across geographic sites. 
We extended our analysis to handle such spatial imbalances in the workload. Let the total workload of $\lambda$ req/s be arbitrarily partitioned across the $k$ edge sites with edge site $i$ 
receiving $\lambda_i$ req/s. Across all edge sites $\sum_i \lambda_i=\lambda$.
 The fraction of the total workload seen by edge site $i$ is given by $w_i = \lambda_i/\sum_j\lambda_j$ . Since each edge site sees a different fraction $w_i$ of the requests, the utilization and queuing delays of each edge site will be different. Specifically in Lemma~\ref{lemma:mmk}, the waiting time seen by requests at site $i$ will be 
\begin{equation}
E[w_il w_i>0] = \frac{\sqrt{2}}{1-\rho_{edge_i}} 
\label{eq:wi}
\end{equation}
where $\rho_{edge_i} = \lambda_i/\mu$.

The average queuing delays seen across the edge is the weighted 
average of the queuing delays seen by each site.
That is, $\sum w_i \sqrt{2}/(1-\rho_{edge_i})$
Hence, Lemma 3.1 can be restated as follows. 
\begin{lemma} \label{lemma:Spatial}
In the scenario where the workload sees spatial skews and 
is partitioned unequally across edge site, the edge will offer worse end-to-end latency when 
\begin{equation}
    \Delta n < \sqrt{2}\left( \sum_i \frac{W_i}{1-\rho_{edge_i}} -\frac{1}{\sqrt{k}(1-\rho_{cloud})}\right)
\end{equation}

\end{lemma}

\noindent\textbf{Practical takeaways:}
When the workload exhibits spatial or geographic skews,  Equation \ref{eq:wi} in the above lemma indicates that sites with higher workloads will experience higher queuing delays.  In such cases, the application should no longer deploy an equal number of  servers at each edge site---edge locations that see higher workloads should be provisioned with higher processing capacity (in terms of bigger servers or more servers).  Specifically, each  each edge site should be  allocated server capacity in proportion to the  workload seen by that site.  Further, if the spatial distribution of the workload changes over time, the allocated processing capacity at each site should also be adjusted dynamically to match these workload changes.  So long as the processing capacity at each edge site matches the incoming workload, the edge sites will be balanced in their utilization levels, and condition in Lemma~\ref{lemma:Spatial}  reduces  to  that in Lemma \ref{lemma:mmk}. 
It is important to note that even with the processing capacity of an edge site is matched to the spatial skews in the workload, the performance 
inversion problem does not go away. The edge can still  yield worse end-to-end latency it the condition of Lemma 3.1 is satisfied

\section{Experimental  Comparison}
\label{sec:eval}

Having analytically compared edge and cloud latencies from the perspective of performance inversion, we now experimentally compare the two using realistic applications and real cloud workloads.
\subsection{Experimental Setup}

\noindent \textbf{Edge/Cloud deployments.} To demonstrate that edge performance inversion can occur in real world 
settings, we conduct our experiments on Amazon's EC2 cloud. The EC2 cloud platforms offers numerous choices of cloud data center 
locations which allows us to conduct experiments on cloud servers with different network latencies. 
We utilize these choice of data center locations to experiment with different edge and cloud application scenarios with different network latencies. All our experiments assume an edge round trip latency of 1ms, which represents the best case scenario for edge deployments. We do so by placing the end client (emulated using a workload generator) and the edge servers in two different availability zones (data centers) within the same EC2 availability region, yielding a very low user-to-edge network lantecy.
We use four different cloud deployments to complement our 1ms edge deployment.\\
Our first scenario assumes a \emph{nearby} cloud that is around 15ms away, yielding a small $\Delta n$ ($\Delta n < 20$). As the number of  cloud data centers has grown, it is now feasible for a cloud server to have less than 20m round trip time (RTT) latency from the user.  We place the end-client and the edge servers in the us-east-2 EC2 region (in Ohio), and the cloud servers  in the us-east-1 in Virginia, with an average Round-Trip-Time (RTT) of around 15 ms.~\footnote{We also ran similar experiments with the edge in Ireland and the Cloud in London, but omit these results since they are similar.}\\
Our second scenario assumes a cloud that is between 20 to 30 ms away, yielding a medium  $\Delta n$ ($\Delta n \sim 20$). This is an increasingly common latency seen by end-users in populated urban centers. Our experiments use two configurations. We use an  edge deployment  in Ireland, and a cloud deployment   in Frankfurt, yielding a network RTT between 20 to 24 ms.  We also use an edge deployment in us-east-2 (Ohio),and a cloud deployment in ca-central-1 (Montreal) with an RTT time between 25 to 28 ms.\\
Our third scenario assumes a cloud that is located over 50ms away, yielding a larger $\Delta n$ ($\Delta n > 50$).  Users in many parts of the world may still experience cloud RTTs that exceed 50ms. It is also common to experience higher RTT latencies when using a cellular connection.  In this case, we deploy the edge and end-clients in us-east-2 (Ohio) and the cloud in us-west-1 (North California), with an average RTT of around 50 to 60 ms.\\
Finally, we also experiment with a distant cloud application by deploying the edge  in the US  (us-east-1), and the cloud in Europe (Ireland), with latencies exceeding 80ms. Since this involves crossing trans-continental ocean links, well-designed geo-distributed applications will typically avoid such latencies but we include this scenario for completeness.

All experiments assume the same configuration  for the  cloud and edge servers; we use the c5a.xlarge instance type, with 4 vCPUs, 8 GB of memory, and 10GB/s network inferences. We assume that the cloud deployment of the applications runs on  5 or 10 servers in the cloud, while corresponding edge deployments use one or two servers at each site.  In the cloud case, we use HAProxy, a popular load-balancer to balance the load across the different cloud servers.

\noindent\textbf{Application Workload}  Since many type of edge applications such as mobile AR, visual analytics and IoT processing involve some form or machine learning, we choose deep neural network (DNN) inference workload for our experiments. The application is a web-based  DNN image classification service  built using  Keras, TensorFlow,  and Flask that is designed to receive image requests from an end-client and responds with the class of the image. Note that this is more compute-intensive workload than simple web applications.

We built a workload generator using Gatling, an open-source load- and performance-testing framework based on Scala. We choose Gatling for its scalability. We use a large images dataset from Kaggle. Each second, the workload generator randomly selects a set of images, based on the number of requests configuration for the experiment, and sends them to the edge application for classification. When the response is received, the workload generator logs the end-to-end response time.  The workload generator is capable of generating requests that follow a specified distribution or it can replay a request trace.


\noindent\textbf{Azure Trace Workload} In addition to synthetic workload generation, we use traces from the Azure Public Dataset \cite{shahrad2020serverless} for our evaluation. This dataset provides request traces seen by serverless functions that run on the Azure cloud. To construct the edge site specific workload from these traces, we first choose a set of functions that belong to the same applications and group them into $k$ mutually exclusive sets. The request traces for each grouping of functions is then mapped onto one edge site. Figure \ref{fig:serverless} depicts the 
workload seen by five edge sites that we construct from the aggregate trace and shows spatial and temporal variations that are present in the trace.  The cloud trace is then the aggregated request trace from all edge sites. In addition, the Azure dataset provides execution times of serverless functions as coarse-grain distributions; we sample these distributions to generate an execution time and append it to each request in the trace. When replaying the trace, an image of an appropriate size is chosen to generate a request with the appropriate service time. 

\noindent\textbf{Research questions} Our experimental comparison is designed to answer several research questions: (1) How likely is the edge performance inversion problem in real-world cloud and edge setting under realistic workloads?  (2) What types of edge utilization levels result in a performance inversion? (3) How well do our experimental results validate the predictions of our analytic models? (4) How do tail response times of the edge compare with the cloud? (5) How do cloud locations with different network latencies impact our results?



\subsection{Mean Latency Comparison and  Validation}

First we benchmark our  application while running on the c5a.xlarge EC2 instance type.
Since DNN inference requests are compute intensive,  we find  that  the system reaches 100\%  utilization at 13 req/s where it starts dropping requests or thrashing. Hence our experiments
assume 12 req/s to be the maximum practical sustainable workload on a c5a.xlarge  instance, which
yield utilization levels of around 90\%.

\begin{figure*}
\centering
\begin{minipage}[t]{0.3\textwidth}
\centering
    \includegraphics[width=1\linewidth]{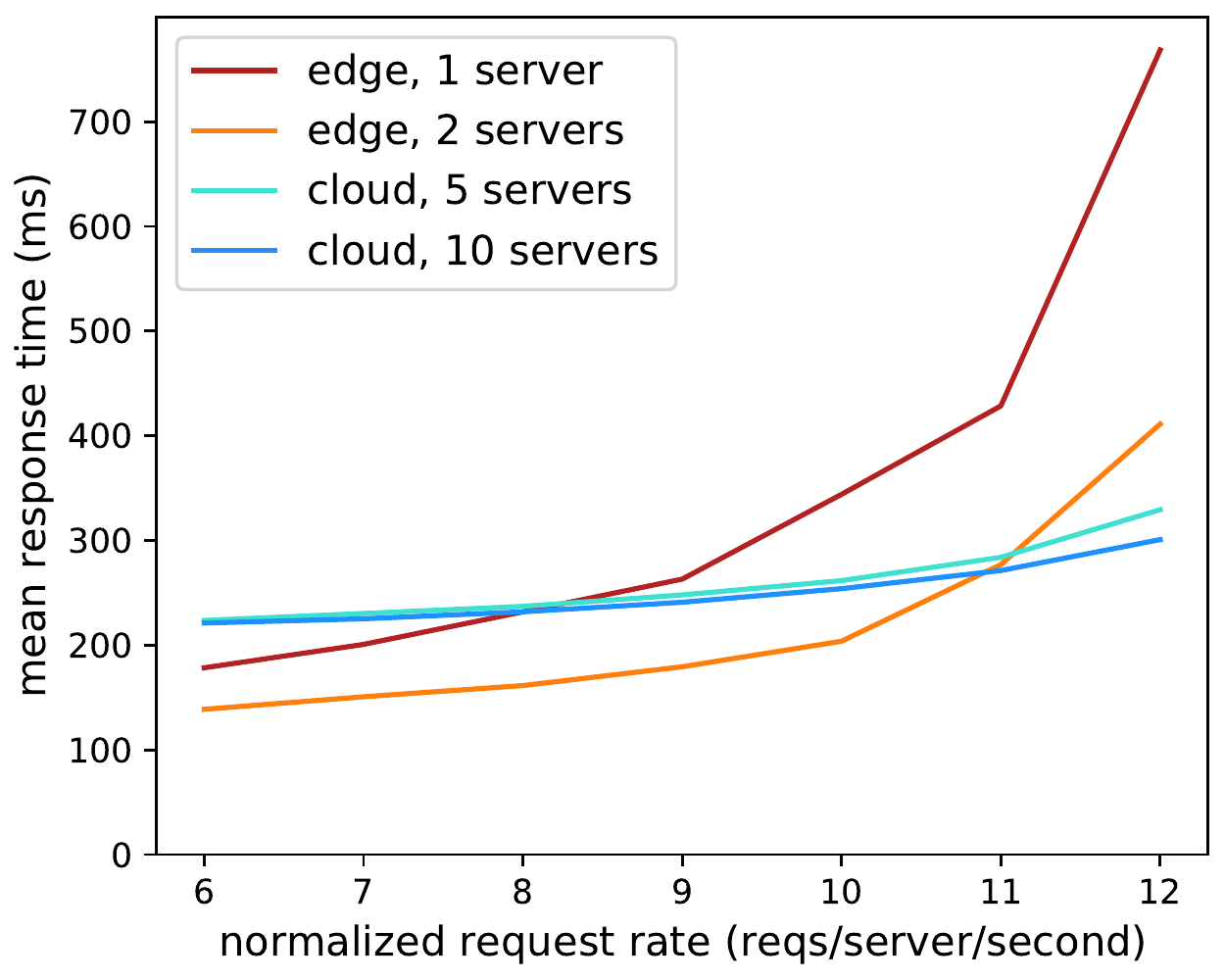}
    \caption{Mean latency of edge (Ireland) and a typical cloud (Frankfurt, \textasciitilde{25ms})} 
    \label{fig:nearCloudl}
\end{minipage} \qquad
\begin{minipage}[t]{0.3\textwidth}
    \centering
    \includegraphics[width=1\linewidth]{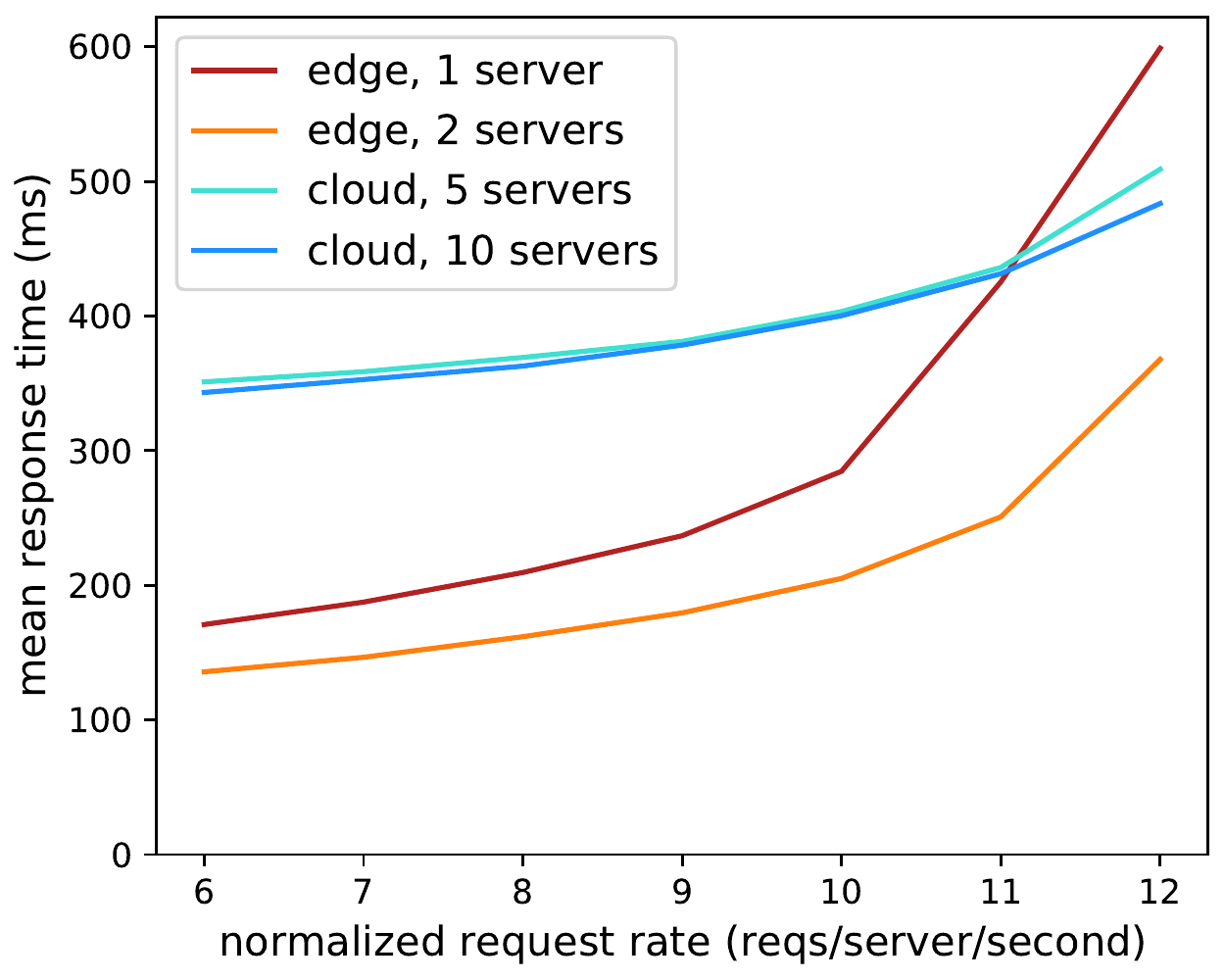}
    \caption{Mean latency of edge (Ohio) and distant cloud (N California, \textasciitilde{54ms})} 
    \label{fig:distantCloud}
\end{minipage} 
\qquad 
\begin{minipage}[t]{0.3\textwidth}
\centering
\includegraphics[width=1\linewidth]{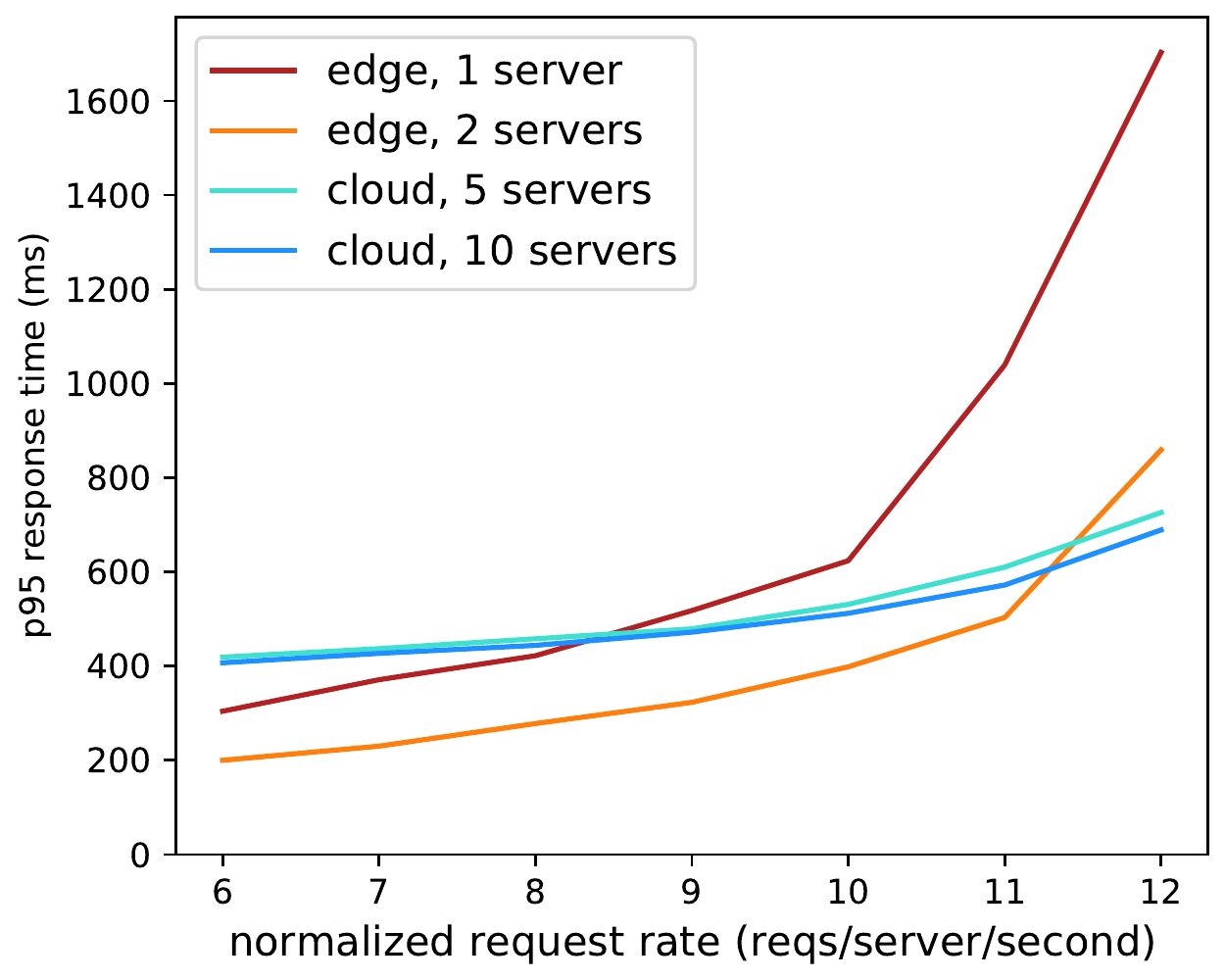}
    \caption{Tail latency of edge (Ohio) and distant cloud (N. California, \textasciitilde{54m}s)}
    \label{fig:p95}
\end{minipage}
\end{figure*}

Our first experiment uses our typical cloud setup with the edge in Ireland and the cloud in the Frankfurt EC2 region. We vary the request rate at each edge server from 6 to 12 req/s and measure the mean end-to-end request latency. We compare this edge setup to a cloud setup with 5 servers that see the cumulative request rate of 5 edge sites (i.e., when each edge server sees $n$ reqs/s, the cloud servers see $5n$ req/s). We also experiment with an edge deployment with 2 servers per edge site and compare it to a cloud deployment with 10 servers. Figure ~\ref{fig:nearCloudl} plots the mean end-to-end latency seen for the edge and the cloud setups. As can be seen, the application provides lower response times from the edge at lower request rates. As the request rate is increased, the edge latency increases faster than the cloud latency and there is crossover point at at 8 req/s where the edge latency becomes higher than the cloud latency, causing a performance inversion beyond this workload level. A similar behavior is seen for the case where the edge has two servers per site and the cloud has 10 servers, but the cross over occurs at a higher workload of around 11 req/s.  

These results show that the performance inversion can occur even in  very low latency edge environments (e.g., 1ms network round trip latency) and at moderate utilization levels of $\rho=8/13 = 0.61$.  Our corollary \ref{corr:equal_balanced} predicts a cutoff utilization of $\rho_{edge}=0.64$ for $\Delta n=30$ and $k=5$, which is within 4.5\% of the experimentally observed value. For the two server edge case and $k=10$, our analytical result predict a cutoff utilization of  $\rho_{edge}=0.75$ , which is within 6\% of the measured value.\footnote{Some of this model error is a  result of the discrete  changes in $\rho$  caused by discrete changes in the request rate as it is varied from 1 to 12 req/s.} Thus, our analytic models can also predict the performance inversion utilization threshold with good accuracy. 

\subsection{Tail Latency Comparison}

There has been considerable research on reducing tail latencies of cloud applications to improve overall end-user experience \cite{li2014tales,suresh2015c3}.  While our analytical results only permit a comparison of mean latencies of the edge and the cloud, we can use experimentation to compare tail latencies as well as the distributions of the edge and cloud latencies.  
To do so, we repeat the previous experiment with an edge deployment in Ohio and the cloud deployment in N. California. Note that unlike the previous experiment where the cloud was 20-30ms away, the edge deployment here is 50-60ms away making a 1ms edge even more beneficial from a network latency perspective. Figure \ref{fig:distantCloud} plots the mean latencies of the edge and the cloud deployments. Since the cloud is more distant, the figure shows that edge latencies are lower than the cloud for a wider range of utilization values.  For the 5 cloud server case, a performance inversion is seen at 11 req/s (i.e., 84.6\% utilization). For the 10 server cloud deployment, we do not see an inversion even at 12 req/s, indicting the inversion occurs close to saturation rate of 13 req/s. Not surprisingly, when the cloud is more distant, edge performance inversion is less likely and will occur at higher utilization level. 

Figure ~\ref{fig:p95} plots the tail latency of the edge the cloud, where tail latency is defined to be the 95-th percentile of the end-to-end latency distribution.  The figure reveals an interesting insight that goes beyond our analytical results. It shows that when tail latencies are concerned, performance inversion occurs a {\em much lower utilization} level than that for the mean latency. Further, even when the edge is well-behaved (i.e., offers lower mean latency than the cloud), it can still see a performance inversion with respect to the tail latency. The figure shows that performance inversion 
occurs at 8 req/s (61\% utilization) for the 5 cloud server case and for the case of 10 cloud servers with 2 servers per edge site, it occurs at 11 req/s (84\% utilization). Note that at these level, the mean edge latency is lower than the cloud latency and there is no inversion with respect to the mean.  Figure \ref{fig:violin_tail} shows violin plots of distributions of the end-to-end latency seen at the edge and the cloud for 10 req/s. As can be seen, edge requests see a higher variability in the end-to-end latency than cloud requests, and the latency distribution at the edge has a longer tail than the cloud. 

\subsection{Impact of Cloud Locations}

\begin{figure*}
\centering
\begin{minipage}[t]{0.45\linewidth}
\centering
\includegraphics[height=2 in]{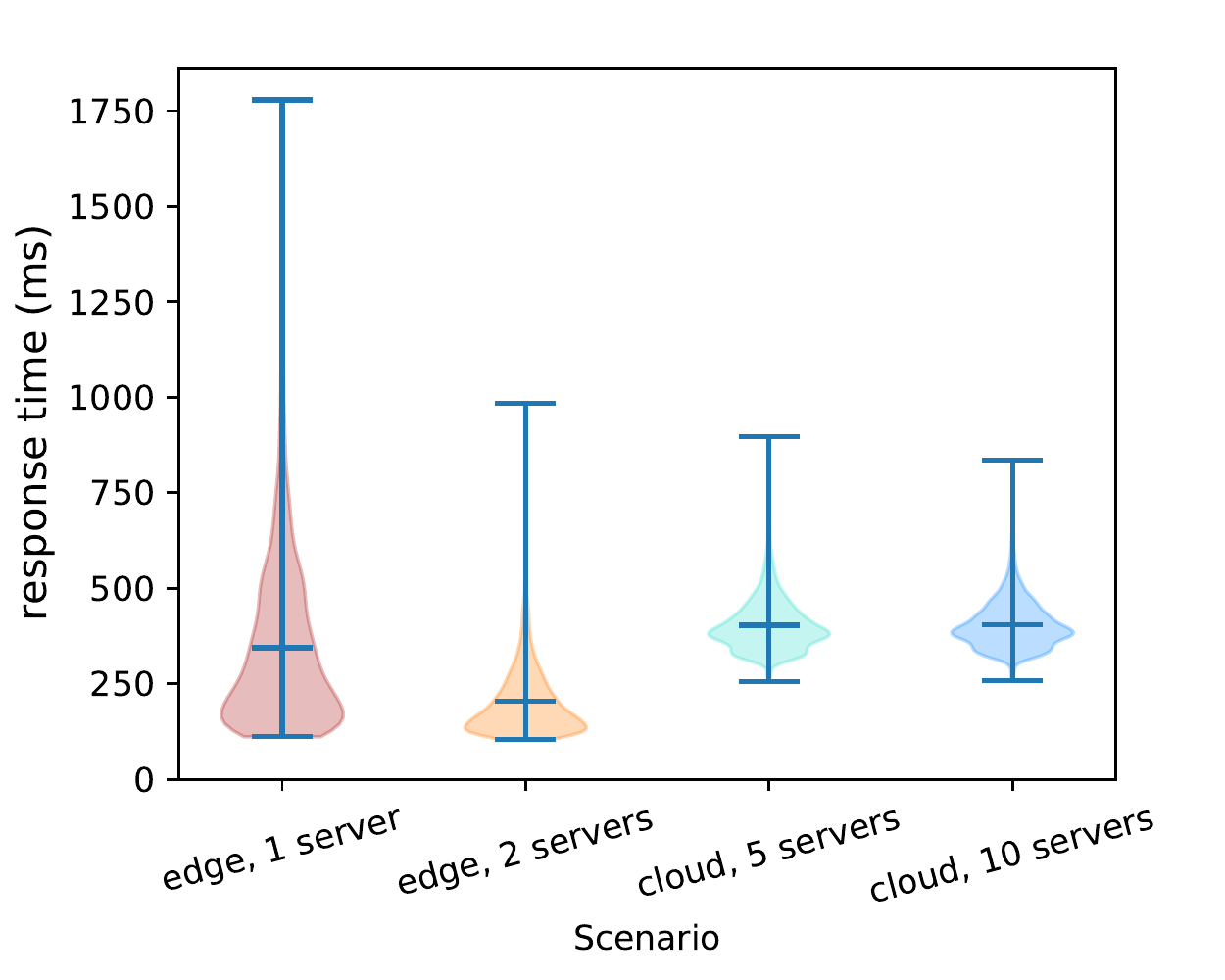}
    \caption{Response distribution of edge (Ohio) and distant cloud (N. California) for 10 req/server/s.}
    \label{fig:violin_tail}
\end{minipage} \qquad
\begin{minipage}[t]{0.45\linewidth}
\centering
    \includegraphics[width=2.5in]{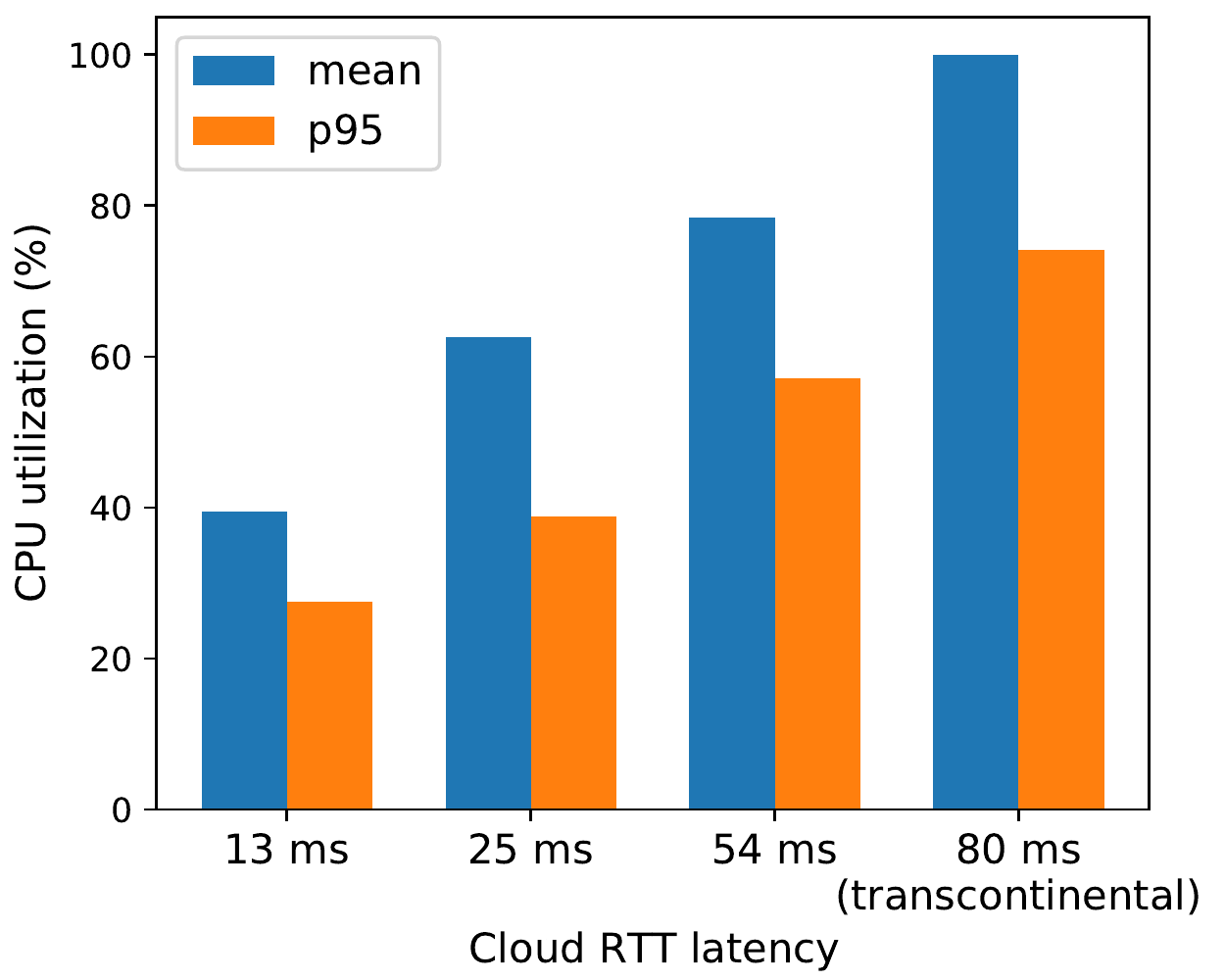}
    \caption{ Utilization levels  above which the edge provides worse mean and tail latencies for various cloud settings. }
    \label{fig:cpu_util}
\end{minipage}
\end{figure*}

We next study the impact of varying the cloud latency of the application by deploying the application at various cloud locations and measure the cutoff utilization where the edge latency becomes worse than the cloud. Figure \ref{fig:cpu_util} plots cutoff utilizations for the mean and tail latency for various cloud deployments (the edge is 1ms away in all cases). The figure shows that as the cloud gets closer to the user, the cutoff utilization for a performance inversion keeps decreasing. For a 15ms cloud (US-east-1), the cutoff utilization for the mean is only 40\% and that for the tail latency is even lower at 25\%. For a 25-30ms cloud, the cutoffs are 60\% and 40\% respectively. For a transcontinental cloud (80ms RTT), the cutoff for the mean is close to saturation but that for the tail is 75\%. The figure shows that as cloud deployments increase, edge performance inversion can occur at progressively lower utilization levels, making it more likely in real-world deployments.

\subsection{Comparison using Azure Workloads}

\begin{figure*}
\centering
\begin{minipage}[t]{0.3\linewidth}
    \centering
    \includegraphics[height=1.6 in]{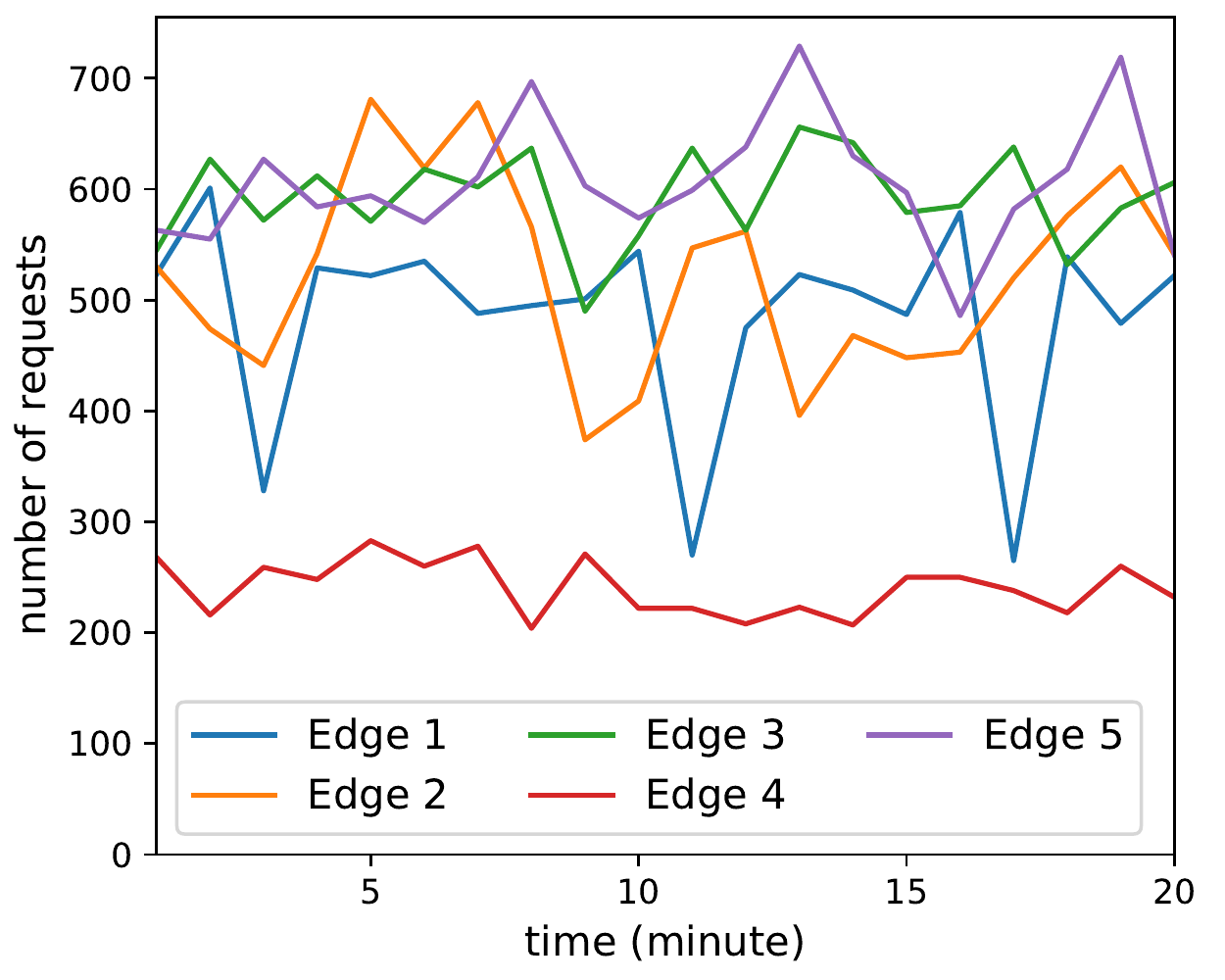}
    \caption{Edge workload of five edge sites based on the Azure serverless traces.}
    \label{fig:serverless}
\end{minipage} \qquad
\begin{minipage}[t]{0.3\linewidth}
    \centering
    \includegraphics[height=1.6 in]{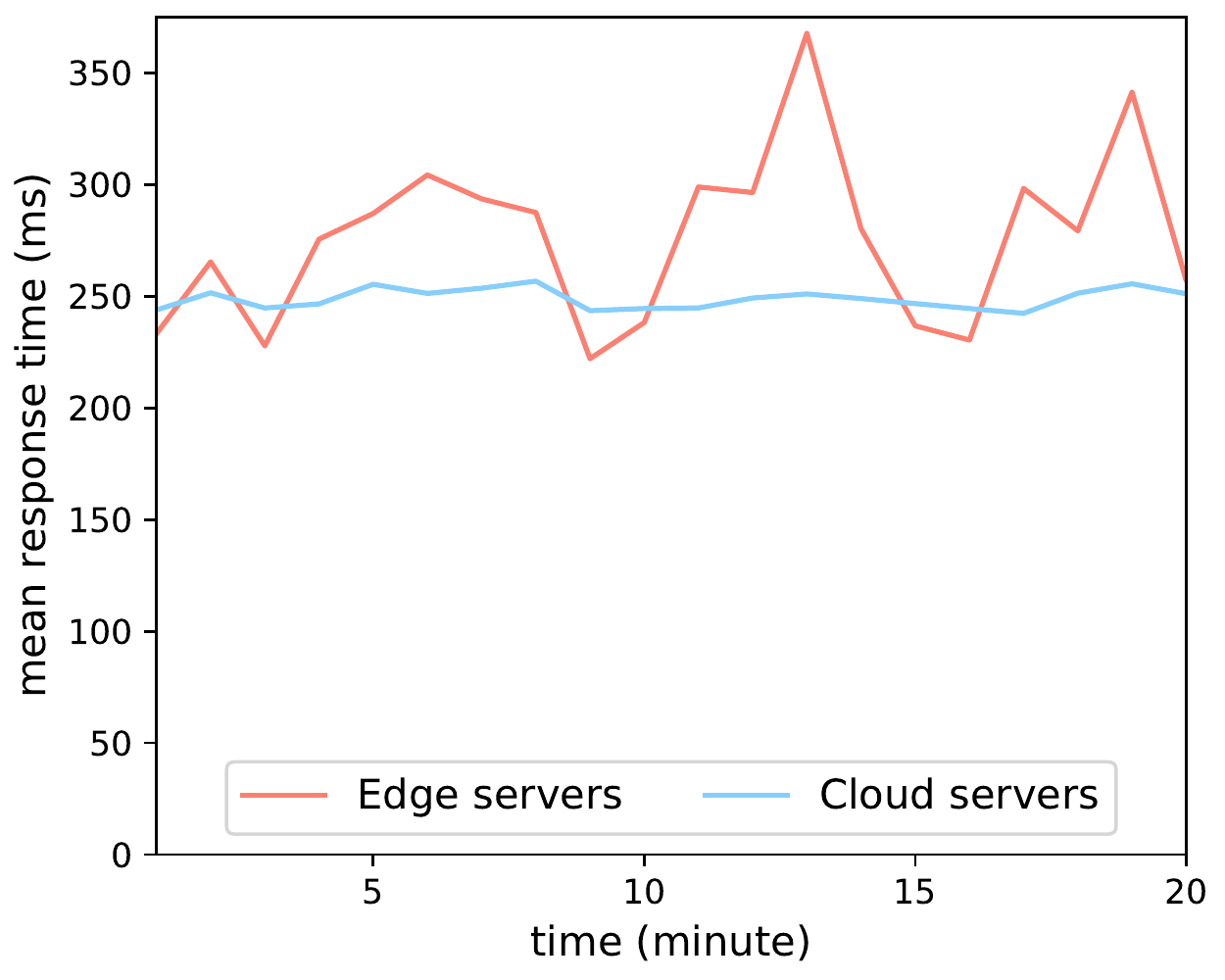}
    \caption{Mean edge and cloud latencies for Azure trace worlkoad.}
    \label{fig:azure}
\end{minipage}\qquad
\begin{minipage}[t]{0.3\linewidth}
    \centering
    \includegraphics[height=1.6 in]{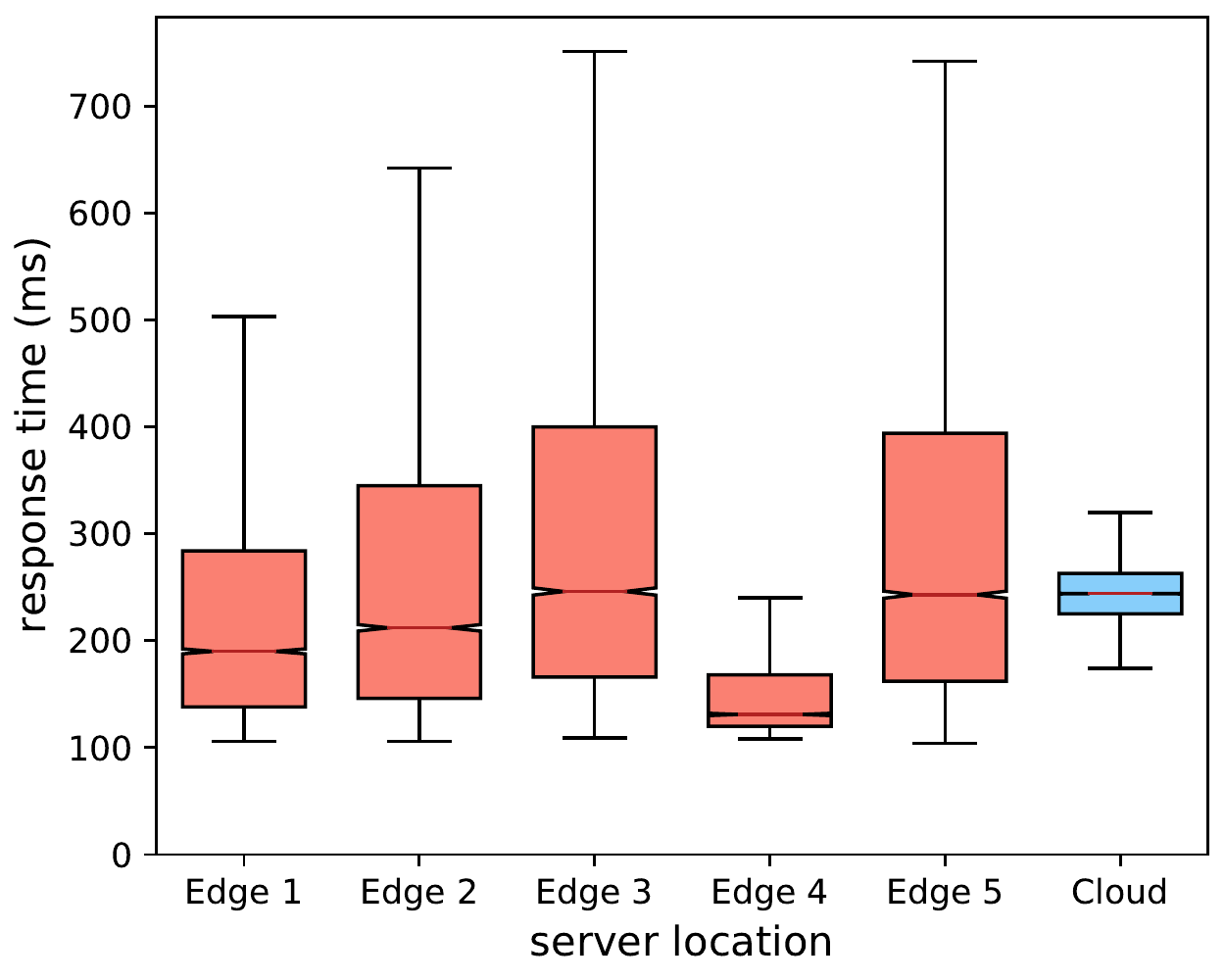}
    \caption{Higher tail latency of the edge under the Azure workload.}
    \label{fig:azure-box}
\end{minipage}
\end{figure*}

Our final experiment involves replaying trace workloads seen at real cloud, i.e., using  Azure public cloud traces.  The experiment emulates generalized distributions for arrival rates and service times and also includes  dynamic spatial and temporal workload skews, as shown in Figure \ref{fig:serverless}.
We replay the trace workload shown in Figure \ref{fig:serverless} at each of the five edge sites and the cumulative workload to the cloud site with 5 servers. We use the typical case cloud scenario with the edge at US-east-2 (Ohio) and the cloud in Montreal (ca-central-1), with RTT of 25 to 28ms. Figure 
\ref{fig:azure} compares the mean latency seen by requests across all edge sites to the mean cloud latency. The temporal fluctuations across sites causes the utilization of each edge site to vary over time and edge sites frequently see a performance inversion, causing the mean edge latency to rise above the cloud latency. The cumulative workload at the cloud sees a smoothing benefits and there is less variation in the cloud latency as a result.  Figure \ref{fig:azure-box} shows a box plot of the latency seen by each of the five edge sites and the cloud. The figure show the unequal partitioning of the workload across edge sites causes different edge sites to exhibit different latencies; the more bursty the workload, the more variable the latency distribution. Edge site 4 see a lower workload than the rest and also offers the lowest latencies as a result. 


\noindent{\bf Practical takeaways} Our experimental results confirm our analytical results and show that edge performance inversion can occur in real-world edge and cloud deployments. The edge can experience a performance inversion even at moderate utilization level in typical deployments. Even when the edge is well-behaved with respect to mean latency, it can still experience a performance inversion for tail latencies. With increasing workload skew or cloud penetration, edge performance inversion can occur at progressively lower utilizations, making it more likely in real-world edge deployments.

\section{Design Implications}
We now discuss how our results can be used by application developers and cloud platforms to avoid edge performance inversion.

\subsection{Application Considerations} 
An application developer can use our results in three different ways: (i) estimate the chances of a performance inversion for a particular edge deployment, (ii) provision additional capacity at the edge site to eliminate or reduce the chances of a performance inversion, and (iii) employ run-time techniques such as geographic load balancing. 
\noindent\textbf{Estimate Chance of a Performance Inversion}
An application designer can use the edge versus cloud latency of a particular setup to determine
the cutoff utilization for performance inversion to occur (e.g., using  Corollaries~\ref{corr:equal_balanced} and~\ref{corr:large-k}).   Typical cloud application are provisioned for the peak workload, and it is well known that they see low average utilization.  Assuming that edge applications are also provisioned similarly, we should expect edge sites to also have low utilization levels on average. If 
the estimated edge utilization is  lower than the cutoff utilization, then the chance  of edge performance inversion is likely to be small. 

\noindent\textbf{Provision extra capacity:} One approach for avoiding edge performance inversion  is to provision  extra server capacity at each site. In this case, the aggregate  server capacity 
at all edge sites is greater than the $k$ servers of the cloud deployment.  One simple rule of thumb to guide the amount of overprovisioning  at the edge is to use Lemma~\ref{lemma:mmk}. Suppose edge site $i$ has $K_i$ servers and receives $\lambda_i$ req/s. Then for each site $i$,  Lemma~\ref{lemma:mmk} yields 
\begin{equation}
    \Delta n < \sqrt{2}\left( \frac{1}{\sqrt{k_i}(1-\frac{\lambda_i}{\mu k_i})} -\frac{1}{\sqrt{k}(1-\frac{\lambda}{\mu k})}\right)
\end{equation}

Since $\Delta n$, $\lambda_i$, and $\mu$ are all measurable or can be estimated, we have a numerical lower bound on $k_i$ at each site to avoid 
performance inversion. An overprovisioning factor can be applied to $k_i$    to allow sufficient headroom capacity to minimize the probability of performance inversion.

\noindent\textbf{Geographic Load Balancing.} The bank teller analogy of Section~\ref{sec:problem}, which is the basis for edge performance inversion, does not hold if ``queue jockeying’’ ~\cite{ rothkopf1987perspectives} is allowed where a  customer is allowed to switch queues. In the edge case, this amounts 
to request redirection to a different edge site if the local edge site is experiencing high waiting times 
or high utilization. Content delivery networks have 
employed such geographic load balancing methods to avoid overloading a local edge site \cite{nygren2010akamai}. 
Edge performance inversion can be avoided by employing similar geographic load balancing methods, where requested to an overloaded edge sites are redirected to nearby edge sites with space capacity.

\subsection{Edge Cloud Platform Considerations}
The edge performance inversion problem has also important implications 
for edge cloud providers. Each edge site will host applications belonging to multiple customers, each of which see a time varying workload. Like any cloud, an edge cloud has to be provisioned with enough server capacity to serve peak demand across customers. 
This is done by deploying enough servers to serve a high  percentile of the workload across all customers. If the workload arrivals are Poisson, the well known two sigma rule can be used to approximate the 95th 
percentile of the workload. The two sigma rule states that the 95th percentile of  the workload is $\lambda+2\sigma$, where $\sigma$ 
is the standard deviation of the workload. For Poisson 
arrivals, the standard deviation is the square root of the mean, i.e., $\sigma=\sqrt{\lambda}$. Hence, the cloud requires an aggregate server capacity $C_{cloud}=\lambda + 2 \sqrt{\lambda}$ to serve the peak workload. However, for edge sites, and in case of a spatially perfectly balanced workload, each edge site sees  $\lambda/k$ req/s. Hence, each edge site requires enough capacity to serve the peak workload  of $\lambda/k+2\sqrt{\lambda/k}$. Since there are $k$ edge sites, the total edge capacity is
$C_{edge}=k\left( \frac{\lambda}{k}+2\sqrt{\frac{\lambda}{k}}\right)=\lambda+2\sqrt{k\lambda}.$
It follows that  $C_{edge}>C_{Cloud}$ since $\sqrt{\lambda}<\sqrt{k\lambda}$.

This means That even for the simplest type of workload (Poisson arrival, balanced across sites), the peak capacity at the edge is 
higher than 
that  
of the cloud. Intuitively, this is due to the statistical smoothing benefits at the cloud that 
accrue when multiple edge site workload is aggregated at the cloud (which sees a lower peak than the sum of the edge 
peaks) 
To avoid its customers from seeing performance inversion, the degree of overprovisioning at the edge has to be even higher than the above analysis (which did not consider the impact 
of query delays during peak utilization). Our results in Section~\ref{sec:latency-model} only considered the average workload $\lambda$, but not the impact of peak workload $\lambda+2\sqrt{\lambda}$ where queuing delays will be even higher. This implies that cloud providers will incur a higher cost to serve $N$ customers at the edge than the cloud 
and will need to overprovision their edge deployments 
even more than their cloud deployments.

\section{Related Work}

Infrastructures similar to Edge Clouds have been envisioned as early as 1991~\cite{weiser1991computer}, but
were never realized until recently~\cite{satyanarayanan2001pervasive}. 
Satyanarayanan et al. proposed a new architecture that uses a nearby cloudlet, a trusted nearby cluster that mobile devices can use with low latencies to offload some of their computations~\cite{satyanarayanan2009case}. The substantial gains from such an architecture were shown recently, improving response times by 51\% and reducing energy consumption in a mobile device by up to 42\% compared to cloud offloading~\cite{Hu:2016}. However, these gains were not evaluated when the cloudlets are operating under high utilization.

 Research on modeling clouds and data centers has attracted considerable attention~\cite{khazaei2012performance,bruneo2014stochastic,khazaei2013analysis,gandhi2010optimality,gandhi2013exact}. This has formed the basis for some more recent work to analytically study edge clouds, e.g., to decide where and when services should be migrated in response to user mobility and demand variation~\cite{urgaonkar2015dynamic},  analytical models to compare the performance and utilization between single level and hierarchical designs of the edge clouds~\cite{tong2016hierarchical}, and models to capture the energy consumption trade-offs when offloading the computations or running them locally~\cite{mao2016power}.
Centralized processing has been shown to improve the performance of distributed processing systems significantly~\cite{tsitsiklis2011power,tsitsiklis2012power}. Tsitsiklis and Xu~\cite{tsitsiklis2011power,tsitsiklis2012power} analyze a multi-server model capturing the trade-offs between using centralized and distributed processing using a fluid model approach.  They show that the average queue length in steady state scales as a function of the degree of the fraction of centralized servers $p$ and the traffic intensity, $\lambda$, as $log_{\frac{1}{1-p}}1/(1-\lambda)$ when the traffic intensity approaches 1, which is exponentially smaller than M/M/1 delay scaling. 


To the best of our knowledge, our work is the first to derive closed form bounds from queuing analysis on the cost and latency trade-offs between edge clouds and large-scale clouds, showing how these trade-offs play in realistic scenarios.
\section{Conclusion}
In this paper we presented and studied the edge performance inversion problem where higher edge queuing delays due to resource constraints or workload skews offset the lower network latency of the edge when compared to the cloud.  We 
analytically compared edge and cloud latencies using queuing models and analyzed conditions under which the edge can yield worse performance than the cloud. We conducted an detailed experimental performance comparison of the edge and cloud performance using realistic applications and Azure trace workloads  on  real cloud and edge platforms.   Our experiments showed that the mean and tail latencies seen by edge applications can indeed be worse than the cloud even at moderate utilization levels of 40 to 60\%.   We discussed implications of our results for application designers and cloud service providers and provided insights into how such performance inversion problems can be avoided or mitigated. These insights also point to several directions for future work. We plan to design dynamic edge resource allocation techniques that are robust to performance inversion and can optimize edge tail latencies. We also plan to  study the 
economic costs of edge deployments resulting from the need to deploy extra capacity to prevent performance inversion.

\bibliographystyle{abbrv}
\bibliography{sigproc}  
%
%
\end{document}